\documentclass[preprint,5p,a4paper,11pt]{elsarticle}

\usepackage{lineno,hyperref}
\modulolinenumbers[5]
\usepackage{xcolor}
\usepackage{mathptmx}      % use Times fonts if available on your 
\usepackage[linesnumbered,ruled,vlined]{algorithm2e}
\usepackage{wrapfig}

\journal{}

\begin{document}

\begin{frontmatter}
	
\title{Artificial Intelligence in Software Testing : Impact, Problems, Challenges and Prospect}

\author[1]{Zubair Khaliq}\ead{zikayem@gmail.com}
\author[1]{Sheikh Umar Farooq}\ead{suf.cs@uok.edu.in}
\author[1,2]{Dawood Ashraf Khan}\ead{dawood.khan@uok.edu.in}

\address[1]{University of Kashmir, Srinagar, Jammu and Kashmir, India}
\address[2]{hyke.ai}

\begin{abstract}
\textit{Artificial Intelligence (AI) is making a significant impact in multiple areas like medical, military, industrial, domestic, law, arts as AI is capable to perform several roles such as managing smart factories, driving autonomous vehicles, creating accurate weather forecasts, detecting cancer and personal assistants, etc. Software testing is the process of putting the software to test for some abnormal behaviour of the software. Software testing is a tedious, laborious and most time-consuming process. Automation tools have been developed that help to automate some activities of the testing process to enhance quality and timely delivery. Over time with the inclusion of continuous integration and continuous delivery (CI/CD) pipeline, automation tools are becoming less effective. The testing community is turning to AI to fill the gap as AI is able to check the code for bugs and errors without any human intervention and in a much faster way than humans. In this study, we aim to recognize the impact of AI technologies on various software testing activities or facets in the STLC. Further, the study aims to recognize and explain some of the biggest challenges software testers face while applying AI to testing. The paper also proposes some key contributions of AI in the future to the domain of software testing.}
\end{abstract}

\begin{keyword}
Artificial Intelligence \sep Machine Learning\sep Deep Learning \sep Software Testing \sep Software Testing Activities
\end{keyword}

\end{frontmatter}

%\linenumbers
\section{Introduction}
\label{intro}
With recent progress in automated and digitized data acquisition, efficient machine learning and deep learning algorithms, and high computing infrastructure, Artificial Intelligence (AI) applications are now inflating their footprint in areas that were previously expected to be only the domain of human experts. The AI-powered tools have already made significant progress in various fields, including finance, law, medicine, and even arts. In many respects, AI is radically surpassing human intelligence and is approaching the domain of human creativity and empathy. Examples include AI’s spectacular successes in winning Go \hyperref[R1]{[1]}, chess \hyperref[R2]{[2]}, and other board games with humans, and in surpassing humans on fully defined world puzzles. In the domain of NLP, we witnessed how a powerful language model like GPT3 wrote news articles that people found hard to distinguish from prose written by humans \hyperref[R3]{[3]}. We also witnessed DeepMind’s protein-folding AI solving a 50-year-old grand challenge of biology \hyperref[R4]{[4]}. Over the past few decades, there has been substantial significant growth in the software industry driven by the recent advances in  AI. Artificial Intelligence is gradually changing the landscape of software engineering in general \hyperref[R5]{[5]} and software testing in particular \hyperref[R6]{[6]} both in research and industry as well.\\

In the last two decades, AI has been found to have made a considerable impact on the way we are approaching software testing. Since most of the organizations have turned to automation testing to bridge the gap that exists between the growing complexity of deliverable software and the contraction of the delivery cycle yet the gap is stretching at an alarming pace bringing us closer to a tipping point wherein test automation too will fail for us to deliver quality software on time. AI can help us fill this gap and help us streamline our speeding software delivery process, thereby saving a significant amount of time and effort (and likely a sizeable amount of money too). So far, the use of AI has been very successful in the automation of software testing in some areas. Still, much research remains to be carried out on analyzing, understanding and improving the tested software artefacts in order to learn more and develop better techniques to enable modern software systems. \\

Our goal in this study is to identify software testing activities where AI has made a significant impact and greatly enhanced the process within each activity. We also identify AI techniques that have been mostly applied to the process of software testing. Further, we convey the problems identified by the study that the testing community is facing while implementing AI-based solutions to the testing problems. We also provide some key areas where AI can potentially help the testing community.\\

\section{Background}
\label{back}
 
\textbf{\textit{Artificial Intelligence Overview:}} 
The term artificial intelligence was coined by John McCarthy in 1955 at a conference organized by the Dartmouth Conference. The term was used to refer to all "programming systems in which the machine is simulating some intelligent human behaviour". According to John McCarthy, it is “The science and engineering of making intelligent machines, especially intelligent computer programs” \hyperref[R7]{[7]}.  Here we discuss the main branches of AI that have been mostly applied to software testing.\\

\iffalse The main branches of AI are given in the \hyperref[AI_Branch]{[Figure 1]}.\\

\begin{figure}[!h]
\centering
\includegraphics[scale= 0.90]{figs/Figure1.png}
\caption{Schematic overview of the main branches of artificial intelligence (AI)}
\label{AI_Branch}
\end{figure} \fi

\textit{Artificial Neural Network :} Designing artificial intelligence based on a biological neural network gives birth to an artificial neural network (ANN) \hyperref[R8]{[8]}. Like the biological neural network, the ANN is an interconnection of nodes, analogous to neurons. Each neural network has three critical components: node character, network topology, and learning rules. Node character determines how signals are processed by the node. Network topology determines the ways nodes are organized and connected. Learning rules automatically determine how the weights are initialized and adjusted using weight adjustment schemes. This type of network becomes a computational device, which is able to learn through training, consequently improving its performance.\\

\textit{AI planning :} Research on AI planning can be traced back to the logic theorist program designed by Newell and Simon in the 1960s \hyperref[R9]{[9]}. The task of AI planning is to find a series of effective actions in a given planning domain, to ensure that the initial state in the planning problem can be successfully transferred to the goal state after applying the actions  \hyperref[R10]{[10]}\hyperref[R11]{[11]}.\\ 

\textit{Robotics :} Robotics is a branch of AI, that comprises Electrical Engineering, Mechanical Engineering, and Computer Science for the design, construction, and application of robots. An Intelligent Robot is a physically situated Intelligent Agent containing five major components: textit{effectors, perception, control, communications, and power} \hyperref[R12]{[12]}. Effectors are the peripherals of the robot that help it to move and interact with the environment. Perception is a set of sensors that provide the robot with the capability to sense the environment. Control is analogous to the central nervous system and is capable of computations that allow the robot to maximize its chances of success. Communication is how a robot interacts with other agents like humans use language, gestures, and proxemics to interact with each other. \\

\textit{Machine Learning :} Machine learning can be broadly defined as computational methods using experience to improve performance or to make accurate predictions \hyperref[R13]{[13]}. Here, experience refers to the past information available to the learner, which typically takes the form of electronic data collected and made available for analysis. This data could be in the form of digitized human-labelled training sets, or other types of information obtained via interaction with the environment \hyperref[R13]{[13]} \hyperref[R14]{[14]}. \\

\textit{Natural Language Processing (NLP) :} Natural Language Processing (NLP) refers to the AI method of communicating with an intelligent system using a natural language such as English. Processing of Natural Language is required when we want an intelligent system to perform as per our instructions, when we want to hear decisions from a dialogue-based clinical expert system, etc.\\

\textit{Fuzzy Logic :} Fuzzy logic(FL) is a method of reasoning that resembles human reasoning. The approach of FL imitates the way of decision-making in humans that involves all intermediate possibilities between digital values YES and NO. FL is based on the idea that there is no sharp distinction between the two extremes. FL is a method of reasoning that is applied to make decisions by means of a number of rules which are combined with each other to produce a result. The rules are fuzzy sets, which are used as a basis for decision-making.\\

\textit{Expert Systems :} Expert systems are computer applications developed to solve complex problems in a particular domain, at the level of extraordinary human intelligence and expertise. The common features of expert systems can be summarized as follows: 
\begin{itemize}
\item \textbf{Rules} that define the specific problem are formalized in the form of computer procedures in the programming language. 
\item \textbf{Knowledge Base} in the form of a computerized database, stores the problems and solutions for support in the decision-making process. 
\item \textbf{Inference Engine} which processes and evaluates scenarios.\\
\end{itemize}
The problems posed and solutions found are completely transparent to the user. In simple terms, the system functions as a large, intelligent “computerized brain”. \\

\textbf{\textit{Software Testing Overview:}} Software testing is an investigation conducted to provide stakeholders with information about the quality of the software product or system under test (SUT). Usually, a software development organization expends between 30\% to 40\% of total project effort on testing \hyperref[R15]{[15]} and testing consumes more than 50\% of the total cost of a project \hyperref[R16]{[16]}. A higher-quality software is achieved when SUT is failure-free. A failure is detected when the SUT's external behaviour is different from what is expected of the SUT according to its requirements or some other description of the expected behaviour \hyperref[R17]{[17]}.\\

An important element of the testing activity is the test case. Essentially, a test case specifies in which conditions the SUT must be executed in hopes of finding a failure. When a test case reveals a failure, it is considered successful (or effective)  \hyperref[R18]{[18]}. The test cases are usually derived from either the functional specification, or a design specification, or a requirements specification. A test case specification includes:
\begin{itemize}
	\item The preconditions, which describe the environment and state of the SUT before the test case is executed.
	\item The test steps, which describe the actions that should be performed to execute the test case.
	\item The expected results, which describe the expected results of the executing test case.
	\item The actual results, which describe the results of the executing test case.
\end{itemize}

There are different dimensions under which testing has been studied and implemented and these dimensions define the test adequacy criteria, that is, the criterion that defines what constitutes an adequate test \hyperref[R19]{[19]}. A great number of such criteria have been proposed and investigated and considerable research effort has attempted to provide support for the use of one criterion or another. We discuss test adequacy criteria in the following sections.\\

\textit{1) Testing Types: } Two main types of testing are\\
\begin{itemize}
	\item \textit{Manual Testing :} In manual testing, testers execute test cases manually without the use of tools or scripts. In this type of testing, the tester takes over the role of an end-user and tests the software to identify any unexpected behaviour or bug.
	\item \textit{Automated Testing :}  is a form of software testing that uses software tools to execute predefined tests. The software tools used for automated testing are often called test automation tools or test automation frameworks. It relieves the tester from the burden of executing the test cases however the process of planning and writing test cases in the form of test scripts still needs to be carried out manually. 
\end{itemize}

\textit{2) Testing Techniques: } Three main testing techniques have been identified.\\
\begin{itemize}
	\item \textit{Black-box Testing :} Black-box testing also known as functional testing aims to study the external behaviour of software without dwelling on the internal structure of the software. Black-box testing is based on the inputs and the outputs of the software. 
	\item \textit{White-box Testing :} White-box Testing also known as structural testing, on the other hand, creates test cases based on the SUT implementation. Its purpose is to make sure that all structures (e.g., paths, instructions, and branches) of the SUT are exercised during the execution of the test suite \hyperref[R18]{[18]}. 		\item \textit{Gray-box Testing :} Gray box testing is a testing technique to test a software product or application with partial knowledge of the internal structure of the application. The purpose of grey box testing is to search and identify the defects due to improper code structure or improper use of applications.
	\end{itemize}
	
\textit{3) Testing Phases or Testing Levels: } Testing is performed at all levels of the software development lifecycle including development, release, and production. During development  \textit{unit testing} is carried out to test basic units of software like a method or a class. After unit testing, as the basic units combine to form components further testing is carried out for testing these components to ensure that the integration has not bought any unintended bugs and the component are working as per the specification. The process of testing at the component level is called \textit{Integration testing}. Since different teams work on the code simultaneously and there is much reusable and third-party code that is incorporated in the software a further level of testing known as \textit{system testing} has been identified to test the integrated components from these sources and therefore to test the system as a whole.\\

Most often due to the requirements change or addition of functionality to software or due to maintenance the code changes, which may result in bugs crreping in the code apparently resulting in a failure. To tackle this a technique called \textit{regression testing} is incorporated at all levels of testing. Regression testing is the most cumbersome and time-consuming testing technique as it involves testing the SUT whenever a change is incorporated.\\

Before the release, \textit{requirements testing} ensures that the SUT is performing all the functions according to the requirements that have been pre-defined in the software requirement specification document. \textit{Scenario testing} is carried out before the release where scenarios of the SUT are created and the SUT is tested against these scenarios to look for any unintended behaviour of the SUT. \textit{Performance testing} is a testing measure that evaluates the speed, responsiveness, and stability of a SUT under a workload. \\

At production time \textit{alpha testing} is carried in the development environment wherein the developer acts as the user of the SUT and tries to identify any failure. In this testing technique, the developer actually looks at the SUT from the perspective of the user. \textit{Beta Testing} is the testing of SUT in the user environment. Here the user actually interacts with the SUT and the developer just watches and analyses the SUT for any failure.\\
\section{Impact of AI on Software Testing}
\label{Impact}
The areas in which AI techniques have proved to be useful in software testing research and practice can be characterized by their applications in the software testing life cycle (STLC). From planning to reporting, AI techniques have made a dominant imprint across all the stages of STLC. To study the impact of AI on software testing we have identified testing activities or testing facets for which considerable and significant research has been carried out by applying AI. These testing activities cover most of the STLC. \\
\textit{Test Specification :}
At the beginning of the software testing life cycle, the test cases are written based on the features and requirements of the software. The test cases are written in a checklist type test specification document to ensure that every requirement of the software is tested. It includes the purpose of a specific test, identifies the required inputs and expected results, provides step-by-step procedures for executing the test and outlines the pass/fail criteria for determining acceptance. Below we mention the work of two seminal papers where AI has been applied to this activity.\\

Last and Friedman \hyperref[R21]{[21]} demonstrated the potential use of Info-Fuzzy Networks (IFN) for automated induction of functional requirements from execution data. The induced models of tested software were utilized for recovering missing and incomplete specifications, designing a minimal set of regression tests, and evaluating the correctness of software outputs when testing new, potentially flawed releases of the system.\\
Briand et al. \hyperref[R20]{[20]} proposes a methodology that takes as inputs the test suite (a set of test cases) and test specifications developed using the Category-Partition (CP) strategy. Based on the CP specification, test cases are transformed into abstract test cases which are tuples of pairs (category, choice) associated with an output equivalence class (instead of raw inputs/outputs). C4.5 is then used to learn rules that relate pairs (category, choice), modelling input properties, to output equivalence classes. These rules are in turn analyzed to determine potential improvements of the test suite (e.g., redundant test cases, need for additional test cases) as well as improvements of the CP specification (e.g., need to add a category or choices). \\

\textit{Test Case Refinement: } Test case refinement is a planned activity that is employed by testers to select the most effective test cases for execution consequently reducing the testing cost. We identified two AI techniques applied to this testing activity.\\

Info-Fuzzy Networks (IFN) was used by Last and Kandel \hyperref[R22]{[22]} and Last et al.\hyperref[R23]{[23]} who presented a novel approach to automated reduction of combinatorial black-box tests, based on automated identification of input-output relationships from execution data of the tested program.\\
Singh et al. \hyperref[R24]{[24]} details an approach generating test cases from {\bf Z} specifications for partition testing. The learner receives as input the functional specification in Z. As output, the approach produces a classification tree describing high-level test cases. Then the high-level test cases are further refined by generating a disjunctive normal form for them.\\

\textit{Test Case Generation : } After devising a test adequacy criteria it is the job of testers to formulate a test set that satisfies the test adequacy criteria. Since for complex applications, the job of handcrafting test sets is an unmanageable task most of the testers use automatic test case generation techniques. In the last two decades, there has been considerable growing interest in applying AI to automate test case generation and AI has impacted this testing activity significantly.\\

Prior work in this field started in 1996 where authors used the inductive learning method to generate test cases from a finite set of input-output examples \hyperref[R25]{[25]}. Given a program P and a set of alternative programs P’, the proposed approach yields test cases that are adequate in the sense that they are able to distinguish P from all programs in P’.\\
Xiao et al. \hyperref[R26]{[26]} present an active learning framework for black box software testing. The active learning approach samples input/output pairs from a black box and learns a model of the system’s behaviour. This model is then used to generate new inputs for sampling.\\
Li, H., and Lam, C. P \hyperref[R27]{[27]} proposes an Ant Colony Optimization approach to automatic test sequence generation for state-based software testing. The proposed approach directly uses UML artefacts to automatically generate test sequences to achieve required test coverage.\\
Sant et al.  \hyperref[R28]{[28]} develop methods that use logged user data to build models of a web application. Their approach automatically builds statistical models of user sessions and automatically derive test cases from these models.\\
Paradkar et al \hyperref[R29]{[29]} present an automated approach to generating functional conformance tests for semantic web services. The semantics of the web services have been defined using the Inputs, Outputs, Preconditions, Effects (IOPEs) paradigm. Their techniques allow the generation of test cases that can be executed through GUI or through the direct invocation of web services.\\
Li et al. \hyperref[R30]{[30]} uses a modified AI planner to avoid the occurrence of combinatorial explosion problems with AI Planners. They applied the method on the GUI test case generation, and the main idea was to produce the initial test case from the planner firstly and then propose a way of solution expanding to reinforce the generation.\\
Shen et al. \hyperref[R31]{[31]} propose an approach by combining the Genetic Algorithm(GA)  with the Tabu Search technique. The experimental study that they carry out regress over the fact that combining the methods is effective against using the GA method individually for the purpose of generating test cases. The primary observation is that the tabu search helps the proposed technique from getting stuck at some local minimum.\\
Srivastava and Baby \hyperref[R32]{[32]} present an algorithm by applying an ant colony optimization technique, for the generation of optimal and minimal test sequences for behaviour specification of software. The paper presents an approach to generate test sequences in order to obtain complete software coverage.\\
Keyvanpour et al. \hyperref[R33]{[33]} presented some test case generation techniques using memetic algorithms which differentiates from the GA in that at each generation on each individual local optimum is reached using a hill-climbing search. \\
%Thummalapenta et al. \hyperref[R34]{[34]}  created a tool that is designed to lower the cost of initial test automation significantly. Moreover, the tool has the ability to patch scripts automatically for certain types of application or environment changes. In their study they present a detailed case study in the context of a c enterprise web application that has over 6500 manual test cases, comes in two variants, evolves frequently, and needs to be tested on multiple browsers in time-constrained and resource-constrained regression cycles.\\
Verma nad Beg \hyperref[R34]{[34]} propose an approach to generate test cases from software requirements expressed in natural language using natural language processing techniques.\\
Mariani et al. \hyperref[R35]{[35]}present a technique to generate new test cases for GUI-based applications from GUI-driven tests hand-crafted manually. The learner receives as input an initial test suite, GUI actions, and observed states obtained by the tool. As output, this GUI-based testing approach produces a behavioural model from which new test cases can be created.\\
Rijwan and Mohd \hyperref[R36]{[36]} introduce an approach to test case generation for unit software testing by using GA incorporated with mutation analysis. Their algorithm injects mutant into the program and then generate random test cases.\\
Carino and Andrews\hyperref[R37]{[37]} introduce a test sequence generator for GUIs. The system uses ant colony optimization in order to generate tests that have an impact on the state of the GUI.\\
Saddler and Cohen \hyperref[R38]{[38]} expand the notion of goal-based interface testing to generate tests for a variety of goals. They develop a direct test generation technique, EventFlowSlicer, that is more efficient than that used in human performance regression testing, reducing run times by 92.5\% on average for test suites between 9 to 26 steps and 63.1\% across all other test suites.\\
Ansari et.al \hyperref[R39]{[39]} proposes a system that deals with the automatic generation of test cases from functional requirements using Natural Language Processing (NLP). The proposed system is beneficial as it can automatically analyze the functional requirement from Software Requirement Specification in order to extract test cases for testing.\\
Bozic and Wotawa \hyperref[R40]{[40]} contribute to the application of AI for security testing of web applications, by using automated planning for obtaining test suites to test common vulnerabilities. The planning system generates test cases in the form of a sequence of actions that lead from an initial to a final state.\\
Rosenfield et al. \hyperref[R41]{[41]} uses the structure of the GUI elements as features to a classifier to classify similar GUI's. Later they apply a pre-defined set of test cases for that particular group. The idea stemmed from the fact of reusing test cases for applications with similar structures.\\
Santiago et al. \hyperref[R42]{[42]} uses LSTM to generate test cases using an abstract flow learning mechanism for generation GUI tests. The authors present a novel approach that models how human testers produce test-flows as an application-agnostic abstract sequence problem.\\
Hu et al. \hyperref[R43]{[43]}  generates tests by progressively discovering a SUT's behaviour, which is necessary to handle different SUT designs and synthesize only tests reusable in this SUT. This also demands creating a training hand-crafted testing library from which new and dynamic on the fly test cases can be generated.\\
Moghadam \hyperref[R44]{[44]} has used model-free reinforcement learning to build a self-adaptive autonomous stress testing framework that is able to learn the optimal policy for stress test case generation without having a model of the system under test.\\
\hyperref[R45]{[45]} propose a method that automatically extracts homogeneous test cases that are not dependent on the skills and know-how of the engineer writing the test cases from requirements specification documents\\

%%%%%%%%%%%%%%%%%%%%%%%%%%%%%%%%%%%%%%%%%%%%%%%%%%%%%%%%%%%%%%%%%%%%

\textit{Test Data Generation : } Test Data Generation is a software testing activity or process to create test inputs and data based on logical test cases and test scenarios. It is the quality of test data that determines the testing coverage of a SUT.\\

Preliminary works in applying AI to this activity started in 1995 when Jones et al. \hyperref[R46]{[46]} applied GAs to generate test sets by searching the input domain for test data which ensure that each branch of the code is exercised.\\
Roper \hyperref[R47]{[47]} used the same technique except for setting up a threshold level for the test coverage. \\
Rodrigues et al. \hyperref[R48]{[48]} provided a systematic mapping study regarding the application of GA techniques to Test Data Generation activity. The results showed that genetic algorithms have been successfully applied to simple test data generation, but are rarely used to generate complex test data such as images, videos, sounds, and 3D (three-dimensional) models.\\
Tracey et al. \hyperref[R49]{[49]} provided a method with the aim to develop a generalised framework for automated generation of test-case data to satisfy both black and white box testing of functional properties and also non-functional properties. \\
Souza et al. \hyperref[R50]{[50]}  proposes an automated test data generation approach, using hill-climbing, for strong mutation. The idea is that if automatic test data generation can achieve an acceptable level of mutation score, it has the potential to greatly reduce the involved manual effort.\\
Behjati et al. \hyperref [R51]{[51]}  plan to produce synthetic test data instead of live production data. The authors investigate the use of LSTM a type of Recurrent Neural Network for this purpose.\\
Choi et al. \hyperref[R52]{[52]} introduce a tool that automatically generates sequences of test inputs for Android apps. The learner receives as input sequences of actions extracted from the app’s GUI. The output can be seen as a model representing the GUI of the application under test.\\
Padararu et al. \hyperref[R53]{[53]} presented a tool that is able to generate test data for evaluating programs, having as initial input a corpus of example tests. The corpus is clustered and then a sequence to sequence RNN is used to learn generative models that are able to produce new test data.\\
Sharifipour et al. \hyperref[R54]{[54]} proposes a memetic ant colony optimization algorithm for structural test data generation. By using evolution strategies they improve the search functionality of ants in local moves.\\
Cegin et al. \hyperref[R55]{[55]} proposes Machine learning methods as meta-heuristic approximations modelling the behaviour of programs that are hard to test using traditional approaches, where the path explosion problem does occur and thus could solve the limitations of the current state-of-art approaches.\\
Liu et al. \hyperref[R56]{[56]}  proposes a novel deep learning-based approach to solve the challenges of test data generation. Their approach consists of two phases: In the training phase, the monkey testing procedure is used to learn the testers’ manual inputs and statistically associates them with the contexts, such as the action history and the textbox label; In the prediction phase, the monkey automatically predicts text inputs based on the observed contexts.\\
%%%%%%%%%%%%%%%%%%%%%%%%%%%%%%%%%%%%%%%%%%%%%%%%%%%%%%%%%%%%%%%

\textit{Test Oracle Construction : }  Software testing is the process of verifying the correct behaviour of the SUT as per the requirements. To highlight this when a program is run with a certain input a mechanism is needed to distinguish between the correct and incorrect behaviour of the SUT. This mechanism is known in the testing terminology as the oracle problem \hyperref[R57]{[57]}. Below we mention the AI techniques that have been applied to this problem.\\
Jin et al. \hyperref[R58]{[58]} investigated how ANNs can be used to ease the test oracle problem. The authors conclude that ANN's cannot be used directly on some problems for test oracle construction. More preprocessing and analysis needs to be carried out to apply ANN's directly to a problem.\\
Wang et al. \hyperref[R59]{[59]} were the first to study how ML algorithms can be used to automatically generate test oracles for reactive programs even if the specification is missing. Their approach turns test traces into feature vectors, which are used to train the ML algorithm. The model yielded by the algorithm then acts as a test oracle. \\
Vineeta et al. \hyperref[R60]{[60]} outlines two ML approaches toward implementing test oracles. To predict the expected outputs of the SUT the first approach builds on ANNs and the second one builds on decision trees. The applicability of these approaches was examined through an example using a toy program.\\
Shahamiri et al. \hyperref[R61]{[61]} proposes MultiNetworks Oracles based on artificial neural networks to map the input domain to the output domain. The accuracy of the proposed oracle was up to 98.26\%, and the oracle detected up to 97.7\% of the injected faults.\\
Braga et al. \hyperref[R62]{[62]}  uses historical usage data from an application that goes through a Knowledge Discovery in Database step and is then used for training (using AdaBoostM1 and Incremental Reduced Error Pruning (IREP) ) to generate an oracle suitable for the application under test. \\
Chan et al. \hyperref[R63]{[63]}  developed an approach that trains a classifier using a reference model of the SUT. This supervised ML approach groups test cases into two categories: passed and failure causing. \\
Vanmali et al. \hyperref[R64]{[64]} train an artificial neural network by the backpropagation algorithm on a set of test cases applied to the original version of the system. The trained network is then used as an intelligent oracle for evaluating the correctness of the output produced by new and possibly faulty versions of the software.\\
Agarwal et al. \hyperref[R65]{[65]} studied Info Fuzzy Networks (IFNs) and ANNs to determine the effectiveness of these approaches used to implement test oracles. According to the results of this study, IFNs significantly outperform ANNs in terms of computation time while achieving almost the same fault-defection effectiveness. \\
Tsimpourlas et al. \hyperref[R66]{[66]}  aims at solving the test oracle problem in a scalable and accurate way. The authors use supervised learning over test execution traces. They label a small fraction of the execution traces with their verdict of a pass or fail. Then they use the labelled traces to train a neural network (NN) model to learn to distinguish runtime patterns for passing versus failing executions for a given program.\\

%%%%%%%%%%%%%%%%%%%%%%%%%%%%%%%%%%%%%%%%%%%%%%%%%%%%%%%%%%%%%%%%%%%

\textit{Test Case Prioritization: } Test case prioritization involves arranging the execution of test cases in a particular order - to ensure that multiple test runs are carried out in a variety of ways - so that the test cases that are most likely to expose defects are executed earlier in the testing process. Test cases can also be prioritized by the risk, i.e., the severity of the item being tested or the impact of the risk if it were to occur in the testing process, the importance of a test case, or any other factor. The available tools didn't provide the ability to automatically prioritize test cases paving the path to studying this area. AI has been found to have impacted this testing activity significantly.\\
Spieker et al. \hyperref[R67]{[67]} proposes a technique for automatically learning test case prioritization with the goal to minimize the round-trip time between code commits and developer feedback on failed test cases. The proposed method uses reinforcement learning to prioritize test cases according to their duration, previous last execution and failure history. It does so by working under the guidance of a reward function that rewards error-prone tests higher than the ones that are less likely to find an error.\\
Lenz et al \hyperref[R68]{[68]} introduces an approach that groups test data into similar functional clusters using K-means Clustering, Expectation-Maximization Clustering and incremental conceptual clustering. After this, according to the tester's goals, it uses the C4.5 classifier for the prioritization of test cases. \\
Busjaeger and Xie \hyperref[R69]{[69]}  presents a novel approach that integrates multiple existing techniques via a systematic framework of machine learning to rank tests. The framework takes input code coverage information, text path similarity, test age, failure history, and text content similarity to yield a model for efficient prioritization of test cases.\\
Wang et al. \hyperref[R70]{[70]} uses  ANN to learn the count of times a program segment will be visited in the execution of a test case. Later they use the count estimation for the program segments as a part of the feature vector for a test input and feed the vector to another ANN for test prioritization of the test input.\\
Ozawa et al. \hyperref[R71]{[71]} proposes a statistical method to prioritize software test cases with operational profiles, where the system behaviour is described by a Markov reward model. The authors introduce software code metrics as reward parameters and apply the resulting Markov reward model to the test case prioritization problem.\\
Lachmann et al. \hyperref[R72]{[72]}  uses SVM Rank for test case prioritization for manual system-level regression testing based on supervised machine learning. Their approach considers black-box meta-data, such as test case history, as well as natural language test case descriptions for prioritization. Their experimental results imply that the SVM Rank technique improves the failure detection rate significantly compared to a random order.\\
Lachmann et al. \hyperref[R73]{[73]} takes the work further and evaluates the efficiency of ANN, K-NN, Logistic Regression, and Ensemble Methods in their application to test case prioritization of black-box tests based on natural language artefacts. Their results indicate that logistic regression outperforms the other applied ML algorithms in terms of effectiveness.\\
Nucci et al. \hyperref[R74]{[74]} proposes a Hypervolume-based Genetic Algorithm, namely HGA, to solve the test case prioritization problem when using multiple test coverage criteria. The authors deduct that HGA is more cost-effective and it improves the efficiency of test case prioritization.\\

%%%%%%%%%%%%%%%%%%%%%%%%%%%%%%%%%%%%%%%%%%%%%%%%%%%%%%%%%%%%%%%%%%

\textit{Test Cost Estimation: } Software cost estimation is the process of predicting the effort required to develop a software system. The general practice of software development is that there should be no shortfalls in the estimation of software cost, the earlier the cost estimation the better it is for the team. AI techniques have been found effective for deriving predictions regarding the unseen. \\ 
Zhu et al. \hyperref[R75]{[75]} proposes an experience-based approach for test execution effort estimation. In the approach, they characterize a test suite as a 3-dimensions vector that combines test case number, test execution complexity and its tester together. Based on the test suite execution vector model, they set up an experience database, and then apply support vector machine (SVM) to estimate efforts for given test suite vectors from historical data.\\
Badri et al.\hyperref[R76]{[76]}  evaluated ML algorithms to predict test code size for object-oriented software in terms of test lines of code (TLOC), which is a key indicator of the testing effort. The authors used linear regression, k-NN, Naïve Bayes, C4.5, Random Forest, and Multilayer Perceptron to build the models. According to their results, their approach yields accurate predictions of TLOC.\\
Silva et al. \hyperref[R77]{[77]} studied the application of ANN's and support vector regression to estimate the execution effort of functional tests. After an analysis of the test process, the authors concluded that there are larger impacts on software quality when the effort is underestimated. In order to cope with this, a modified cost function was developed to train an artificial neural network, aiming to bias the predictive model to overestimate, instead of underestimating.\\
Cheatham et al. \hyperref[R78]{[78]} purposes to demonstrate a technique using machine-learning (COBWEB/3) to identify attributes that are important in predicting software testing costs and more specifically software testing time.\\
{\bf Findings} The overall impact of AI techniques on particular software Testing activities can be visualized from \hyperref[Impact]{[Figure 2]}.

\begin{figure}[!h]
%\centering
\includegraphics[scale= 0.45]{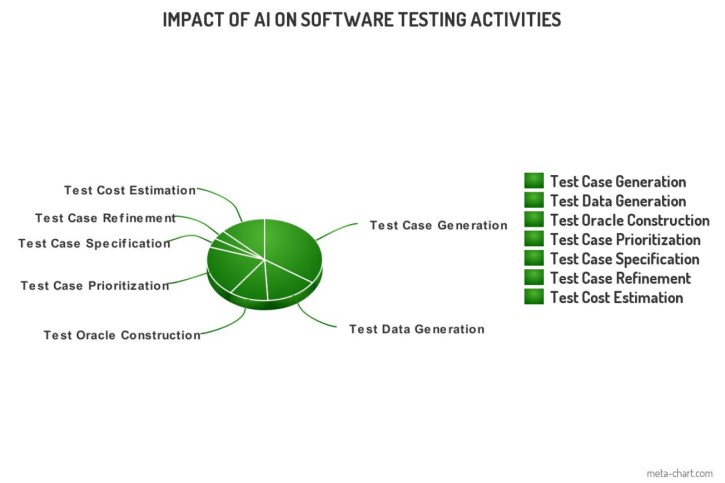}
\caption{Impact of AI on software Testing Activities}
\label{Impact}
\end{figure}

Based on the discovered publications, seven software testing activities viz test case generation, test oracle generation, test data generation, test case prioritization, test case specification, test case refinement, and test cost estimation were identified as activities that have been improved significantly by the application of AI techniques. From this study, we can infer that test case generation or test case design activity has been considerably enhanced by the application of AI techniques. Most of the recent research has been carried out around activities like test case generation, test case prioritization, test data generation and test oracle construction. The trivial reason for this is that these activities are more important than other activities in the STLC. We skipped some software testing activities including test harness, testing technique selection, test repairing, change proness etc. from our study as just one or two AI-based studies have been carried for these activities. \hyperref[table]{[Table 1]} shows a list of AI techniques that have been applied for software testing activities.\\
Also, the most commonly used AI techniques applied to soft testing appear to be solving the problem of optimization across various software testing activities. Specifically, genetic algorithms, ANN, and reinforcement learning were among the techniques that were used across various testing activities more frequently than others.\\

\begin{table*}[!]
\label{table}
\caption{AI techniques applied to software testing activities.}
\begin{center}

%\centering
 \begin{tabular}{p{5cm}p{12cm}} 
\textbf{Software Testing Activity} & \textbf{AI Technique Applied} \\
 \hline
Test Case Generation & Inductive Learning - Active Learning - Ant colony Optimization - Markov Model - AI Planner -GA - Tabu Search - NLP - Re-enforcement Learning - C4.5 - Goal Based - Decision Tree - K-Neirest Neighbour - Logistic Regression - Random Forest - Multi-Layer Perceptron - K star - LSTM - Heuristic Search \\
\hline
Test Data Generation&GA - Simulated Annealing - Hill Climbing - Generative Model - LSTM - Deep Re-enforcement Learning - Ant Colony Optimization - Heuristic Methods\\
\hline
Test Oracle Construction&ANN - SVM - Decision Trees - AdaBoostM1  - Incremental Reduced Error Pruning (IREP)  - Info Fuzzy Network\\
\hline
Test case prioritization&K-Means - Expectation-Maximization - c4.5 - Cob Web  - Reinforcement Learning - CBR - ANN - Markov Model - K-NN - Logistic Regression - SVM Rank\\
\hline
Test Case Specification&IFN - C4.5 \\
\hline
Test Case Refinement&IFN - Classification Tree Method \\
\hline
Test Cost Estimation&SVM - linear regression - k-NN - Naïve Bayes - C4.5 - Random Forest - Multilayer Perceptron \\
\hline
\end{tabular}
\end{center}

\end{table*}

\section{Problems and Challenges of AI in Software Testing}
\label{Problems}
Considering the lack of industrial expertise and research work this section outlines some of the open problems and challenges in the application of AI to software testing.\\

\textit{Test Oracle the biggest challenge :} Test oracle problem is a companion of every researcher and practitioner working in the field of software testing. It has been there from the inception of the software testing conundrum and the problem is expected to stay with us for a longer duration or maybe forever. Despite continuous attempts to mitigate the problem of the test oracle, researchers have been able to solve this problem for a static subset of SUT's. As soon as the dynamic traits of the SUT start to display the previous test oracle derived for the SUT starts to lose effectiveness. In many scenarios, even the documentation from which test oracles are generated is missing in the requirements document.  To cope with this dynamism and to dream of a documentation-free effective test oracle AI techniques have been employed and these AI techniques have provided a significant initial effort towards realizing this dream.\\

\textit{Availability of Data :} Any model in AI must be trained and tested before being deployed in production. The efficiency and effectiveness of a model are highly correlated to the amount of data with which a model is trained and tested. Acquiring data for building AI models in the domain of software testing is a challenging task because software testing unlike other fields of study is not fully automated yet. Apparently, a lot of testing is still being carried manually and it is difficult to capture data when testing is manual. This exhumes as a bottleneck to data acquisition for training AI models in the future.\\

\textit{Adaptiveness to data :} AI models are highly dependant on the data with which they are trained and tested. An important phase in the production of an AI model is the collection of robust datasets from real-world scenarios and the use of that data to train a model generalized to fit that data. Such a model assumes future data and historic data (data with which the model is trained) to be from the same distribution. However, it is often not the case as most data has higher disparity over time e.g. learning customers shopping behaviour is dependant on seasons. AI models are evaluated for generalizations by testing the model on a particular subset from the data (test set) which is from the same time distribution. Over time less promising outcomes from such models are witnessed. Some AI techniques allow the model to be readjusted to adapt to the changes in the new data. However, the challenging task is to detect this ideal time to readjust and even automate the readjustment process.\\

\textit{Identifying Test Data :} Every AI model must be tested thoroughly before being put to production. Model testing is like a black-box technique where the structural or logical information regarding the model is not a necessity. Rather comprehensive information and understanding regarding the testing data is required. Again the selection of testing data from the same distribution can incur issues resulting in a biased model. The problem is with the coverage of the test set   i:e asking the question "Is the model tested over a larger distribution of data?" Identification of such coverage-based test datasets is a challenging task in the domain of software testing.\\

\textit{Exhaustive search space leads to generality loss :} For most of the optimization problems in search-based software testing, the AI algorithm has to exhaustively search for the solution or the goal. Although sub-optimal search strategies have been identified and implemented so far they work for a particular class of problems. To include more general solutions to a variety of problems the inputs of the whole problem domain are to be included, consequently making the input space more exhaustive. Versatile and broadly capable AI methodologies need to be identified to cope with this generality loss.\\

\textit{Exploitation of Multicore Computation :} A lot of AI techniques are highly computationally expensive making them potentially incompatible with large-scale problems faced by software testers. With the recent advancements in computing infrastructure, Graphical Processing Units (GPU) and Tensor Processing Units (TPU) have been incorporated at scale for these techniques. More work is required to fully exploit the enormous potential of the rapidly increasing number of processors available. Since such high computational devices are expensive more work needs to be carried out towards designing techniques that require less computation and still match the performance of high computing devices.\\
\section{Prospects of AI in Software Testing}
\label{Prospects}
In the past few years, many companies have begun to invest in AI-powered software testing technologies. These AI systems offer an alternative to traditional testing processes. While AI systems are still relatively new, the potential gains are simply too great to ignore. Here are some excerpts from our study and from software testing industry experts where we expect these technologies to potentially help software testers in the future:
\begin{itemize}
	\item Collaborating with people who are geographically spread out can be difficult. This is where AI systems can be relied upon to carry out routine, labour-intensive tasks. This frees up more productive time for software testers to spend on addressing the most complicated issues.
	\item Simulated testing - The ability to program AI systems to test application code is incredibly useful. It offers a realistic simulation of a situation that a software tester might face. This also improves the accuracy of tests because they can identify and replicate all possible scenarios.
	\item The next generation of artificial intelligence in software testing will include self-modifying tools that can instantly identify and fix vulnerabilities without any human intervention creating self-healing systems.
	\item With artificial intelligence in software testing, software companies, and testers can reduce their costs by a great degree, which is already happening. We think it will normal to see organizations and other user groups automate their testing process using AI while testers focus on the exploratory testing of systems.
	\item The AI predictive analytics will play a major role in discovering all possible test cases and will make the software products more robust, reliable and will exceed customer expectations.
	\item AI is operating at all levels of testing from planning to execution to reporting and it is expected to take over all the tasks in the STLC which require human intelligence. This in turn will free the tester from the job of various time-consuming testing strategies like regression testing and smoke testing etc.
	\item AI incorporated in testing will provide a high ROI because these systems ensure that the time allocated to deliver the product is spent on polishing its features rather than on testing and debugging technical defects.
	\item AI-powered automation tools will help to increase the level of autonomy in software testing and hence help to deliver higher quality software. AI-related technologies are helping to bridge the gap between human and machine-driven testing capabilities.
	\item AI is expected to impact testing in all the software product areas including mobile applications, web applications, IoT, embedded systems, database applications, gaming industry, real-time applications, critical software applications to name a few. 
	\item With more data being acquired and stored, AI can enhance the software testing capabilities which are somewhat restricted today due to the non-availability of data.

\end{itemize}
\section{Conclusion}
\label{conclusion}
In the last two decades, the rapid growth of interest in topics where AI has been applied to software testing is a testimony to the appetite the software testing community has for AI. This is a consequence of AI providing efficient solutions to the problems faced by the testing community for a long time. AI has already been accepted as a promising solution to many problems faced by testers all around the globe.
In this paper, we studied the impact of AI across all stages of the STLC. We identified seven software testing activities that were most enhanced by AI techniques. GA's, Reinforcement Learning, and ANN were among the most widely used techniques from the domain of AI. We identified problems and challenges researchers and testers face while applying AI techniques to software testing. We also provided a future prospect into how AI can shape the software testing domain.


\begin{thebibliography}{37}
	\bibliographystyle{alpha}
	%Sri Parameswaran, Tilman Wolf, “Embedded systems security - an overview”, Design Autom. for Emb. Sys. 2008.
	\bibitem{RefJ}
	C. Koch, “How the computer beat the Go master,” Scientific American, vol. 19, 2016.
	\label{R1}
	\bibitem{RefJ}
	F.-H. Hsu, Behind Deep Blue: Building the Computer that Defeated the World Chess Champion. Princeton, NJ: Princeton Univ. Press, 2004
	\label{R2}
	\bibitem{RefJ}
	Brown, T. B. et al. Preprint at https://arxiv.org/abs/2005.14165 (2020).
	\label{R3}
	\bibitem{RefJ}
	https://www.technologyreview.com/2020/11/30/1012712/deepmind-protein-folding-ai-solved-biology-science-drugs-disease/
	\label{R4}
	\bibitem{RefJ}
	M. Harman. The role of artificial intelligence in software engineering. In 1st International Workshop on Realizing Artificial Intelligence Synergies in Software Engineering (RAISE 2012), Zurich, Switzerland, 2012.
	\label{R5}
	\bibitem{RefJ}
	Hourani, H., Hammad, A., Lafi, M. (2019, April). The Impact of Artificial Intelligence on Software Testing. In 2019 IEEE Jordan International Joint Conference on Electrical Engineering and Information Technology (JEEIT) (pp. 565-570). IEEE.
	\label{R6}
	\bibitem{RefJ}
	 J. McCarthy, “Programs with common sense,” in Proceedings of the Symposium on Mechanisation of Thought Processes, vol. 1. London: Her Majesty’s Stationery Office, 1958, pp. 77–84.
	\label{R7}
	\bibitem{RefJ}
	Zou J., Han Y., So SS. (2008) Overview of Artificial Neural Networks. In: Livingstone D.J. (eds) Artificial Neural Networks. Methods in Molecular Biology™, vol 458. Humana Press. https://doi.org/10.1007/978-1-60327-101-12
	\label{R8}
	\bibitem{RefJ}
	Newell, A., and Simon, H. A. 1963. GPS: A Program That Simulates Human Thought. In Computers and Thought, eds. E. A. Feigenbaum and J. Feldman. New York: McGraw-Hill. [GPS]
	\label{R9}
	\bibitem{RefJ}
	Jiao, Z.; Yao, P.; Zhang, J.; Wan, L.; Wang, X. Capability Construction of C4ISR Based on AI Planning. IEEE Access 2019, 7, 31997–32008.
	\label{R10}
	\bibitem{RefJ}
	James Hendler, Austin Tate, and Mark Drummond, “AI Planning: Systems and Techniques,” AI Magazine Volume 11 Number 2 (1990)	
	\label{R11}
	\bibitem{RefJ}
	Robin R Murphy, "Introduction to AI Robotics Second Edition, " The MIT Press, Cambridge, Massachusetts, London England, 2019.
	\label{R12}
	\bibitem{RefJ}
	M. Mohri, A. Rostamizadeh, and A. Talwalkar, Foundations of Machine Learning. Cambridge, MA, USA: MIT Press, 2012.
	\label{R13}
	\bibitem{RefJ}
	P. Louridas and C. Ebert, “Machine learning,” IEEE Softw., vol. 33, no. 5, pp. 110–115, Sep./Oct. 2016.
	\label{R14}
	\bibitem{RefJ}
	Pressman, R. S. "Software engineering: A practitioner's approach," New York: McGraw-Hill. 1987	
	\label{R15}
	\bibitem{RefJ}
	Ramler, R., \& Wolfmaier, K, "Economic perspectives in test automation: balancing automated and manual testing with opportunity cost. In Proceedings of the 2006 international workshop on Automation of software test", (pp. 85-91). ACM, 2006, May
	\label{R16}
	\bibitem{RefJ}
	P. Ammann and J. Offutt, "Introduction to Software Testing, 2nd ed,". Cam- bridge, U.K.: Cambridge Univ. Press, 2016.997
	\label{R17}
	\bibitem{RefJ}
	Vinicius H. S. Durelli , Rafael S. Durelli , Simone S. Borges, Andre T. Endo, Marcelo M. Eler , Diego R. C. Dias , and Marcelo P. Guimaraes, "Machine Learning Applied to Software Testing: A Systematic Mapping Study,". IEEE TRANSACTIONS ON RELIABILITY. 2019.
	\label{R18}
	\bibitem{RefJ}
	H. Zhu, P. A. V. Hall, and J. H. R. May, “Software unit test coverage and adequacy,” ACM Comput. Surveys, vol. 29, no. 4, pp. 366–427, 1997.
	\label{R19}
	\bibitem{RefJ}
	INDI,2014.
	L. C. Briand, Y. Labiche, and Z. Bawar, “Using Machine Learning to Refine Black-Box Test Specifications and Test Suites,” 2008 The Eighth International Conference on Quality Software, 2008.
	\label{R20}
	\bibitem{RefJ}
	Mark Last and Menahem Friedman, "BLACK-BOX TESTING WITH INFO-FUZZY NETWORKS," Artificial Intelligence Methods in Software Testing, pp. 1-20 (2004)
	\label{R21}
	\bibitem{RefJ}
	Last M., Kandel A. (2003) Automated Test Reduction Using an Info-Fuzzy Network. In: Khoshgoftaar T.M. (eds) Software Engineering with Computational Intelligence. The Springer International Series in Engineering and Computer Science, vol 731. Springer, Boston, MA. https://doi.org/10.1007/978-1-4615-0429-0\_9
	\label{R22}
	\bibitem{RefJ}
	Last M., Friedman M., Kandel A. "USING DATA MINING FOR AUTOMATED SOFTWARE TESTING," International Journal of Software Engineering and Knowledge Engineering. Vol. 14, No. 04, pp. 369-393 (2004)
	\label{R23}
	\bibitem{RefJ}
	H. Singh, M. Conrad, and S. Sadeghipour, “Test case design based on Z and the classification-tree method,” in Proc. IEEE Int. Conf. Formal Eng. Methods, 1997, pp. 81–90.
	\label{R24}
	\bibitem{RefJ}
	F. Bergadano and D. Gunetti, “Testing by means of inductive program learning,” ACM Trans. Softw. Eng. Methodol., vol. 5, no. 2, pp. 119–145, 1996.
	\label{R25}
	\bibitem{RefJ}
	G. Xiao, F. Southey, R. C. Holte, and D. Wilkinson, “Software testing by active learning for commercial games,” in Proc. 20th Nat. Conf. Artif. Intell., 2005, vol. 2, pp. 898–903.
	\label{R26}
	\bibitem{RefJ}
	Li, H., \& Lam, C. P. (2005). An ant colony optimization approach to test sequence generation for state-based software testing. QSIC 2005, 255–262.
	\label{R27}
	\bibitem{RefJ}
	J. Sant, A. Souter, and L. Greenwald, “An exploration of statistical models for automated test case generation,” in Proc. Int. Workshop Dyn. Anal., 2005, pp. 1–7.
	\label{R28}
	\bibitem{RefJ}
	Paradkar, A. M., Sinha, A., Williams, C., Johnson, R. D., Outterson, S., Shriver, C., \& Liang, C. (2007). Automated functional conformance test generation for semantic web services. ICWS 2007, 110–117.
	\label{R29}
	\bibitem{RefJ}
	Li, L., Wang, D., Shen, X., \& Yang, M. (2009). A method for combinatorial explosion avoidance of AI planner and the application on test case generation. CiSE 2009, 1–4.
	\label{R30}
	\bibitem{RefJ}
	Shen, X., Wang, Q., Wang, P., \& Zhou, B. (2009). Automatic generation of test case based on GATS algorithm. GrC 2009, 496–500.
	\label{R31}
	\bibitem{RefJ}
	Srivastava, P. R., \& Baby, K. (2010). Automated software testing using metahurestic technique based on an Ant Colony Optimization. ISED 2010, 235–240.
	\label{R32}
	\bibitem{RefJ}
	M.R. Keyvanpour, H. Homayouni and Hasein Shirazee, 2011. Automatic Software Test Case Generation. Journal of Software Engineering, 5: 91-101.
	\label{R33}

	
	\bibitem{RefJ}
	R. P. Verma and M. R. Beg, “Generation of Test Cases from Software Requirements Using Natural Language Processing,” 2013 6th International Conference on Emerging Trends in Engineering and Technology, 2013.
	\label{R34}
\bibitem{RefJ}
	L. Mariani, M. Pezze, O. Riganelli, and M. Santoro, “Automatic testing of GUI-based applications,” Softw. Testing, Verification Reliab., vol. 24, no. 5, pp. 341–366, 2014.
	\label{R35}
	\bibitem{RefJ}
	Rizwan Khan and Mohd Amjad, " Automatic Test case generation for unit software testing using genetic algorithm and mutation testing," 2015 IEEE UP section conference on electrical computer and electronics (UPCON), 2015
	\label{R36}
	\bibitem{RefJ}
	Carino, S., \& Andrews, J. H. (2015). Dynamically Testing GUIs Using Ant Colony Optimization. ASE 2015, 138–148.
	\label{R37}
	\bibitem{RefJ}
	Jonathan Saddler, Myra B. Cohen, "EventFlowSlicer: goal based test generation for graphical user interfaces"Proceedings of the 7th International Workshop on Automating Test Case Design, Selection, and EvaluationNovember 2016 Pages 8–15https://doi.org/10.1145/2994291.2994293
 	\label{R38}

	\bibitem{RefJ}
	A. Ansari, M. B. Shagufta, A. S. Fatima, and S. Tehreem, “Constructing Test cases using Natural Language Processing,” 2017 Third International Conference on Advances in Electrical, Electronics, Information, Communication and Bio-Informatics (AEEICB), 2017.
	\label{R39}
	\bibitem{RefJ}
	Bozic, J., \& Wotawa, F. (2018). Planning-based security testing of web applications. AST@ICSE 201, 20-26.
	\label{R40}
	\bibitem{RefJ}
	Rosenfeld, A., Kardashov, O., \& Zang, O. (2018). Automation of Android Applications Functional Testing Using Machine Learning Activities Classification. MOBILESoft@ICSE 2018, 122–132.
	\label{R41}
	\bibitem{RefJ}
	Santiago, D., Clarke, P. J., Alt, P., \& King, T. M. (2018). Abstract flow learning for web application test generation. A-TEST@ESEC/SIGSOFT FSE 2018, 49–55.
	\label{R42}
	\bibitem{RefJ}
	Hu, G., Zhu, L., \& Yang, J. (2018). AppFlow: using machine learning to synthesize robust, reusable UI tests. ESEC/SIGSOFT FSE 2018, 269–282.
	\label{R43}
	\bibitem{RefJ}
	Moghadam, M. H. (2019). Machine Learning-assisted Performance Testing. ESEC/SIGSOFT FSE 2019, 1187–1189.
	\label{R44}
	\bibitem{RefJ}
	Kazuhiro Kikuma, Takeshi Yamada, Koki Sato, Kiyoshi Ueda,"Preparation Method in Automated Test Case Generation using Machine Learning,"Proceedings of the Tenth International Symposium on Information and Communication Technology, December 2019, Pages 393–398, https://doi.org/10.1145/3368926.3369679
	\label{R45}
	\bibitem{RefJ}
	B.Jones,H.Sthamer,X.Yang,D.Eyres,The automatic generation of software test data sets using adaptive search techniques, Proceedings of the Third International Conference on Software Quality Management SQM'95 ,Sevilla, Spain 1995
	\label{R46}
	\bibitem{RefJ}
	M.Roper,Computer-Aided Software Testing using Genetic Algorithms, Proceedings of the 10th International Software Quality Week QW '97), San Francisco, USA 1997). 
	\label{R47}
	\bibitem{RefJ}
	Rodrigues, D. et al. “Using Genetic Algorithms in Test Data Generation.” ACM Computing Surveys (CSUR) 51 (2018): 1 - 23.
	\label{R48}
	\bibitem{RefJ}
	N.Tracey,J.Clark,K.Mander,Automated program flaw finding us- ing simulated annealing, Proceedings of the ACM/SIGSOFT Inter- national Symposium on Software Testing and Analysis IS- STA ’98, Clearwater Beach, Florida, USA 1998
	\label{R49}
	\bibitem{RefJ}
	Francisco Carlos M. Souza, Mike Papadakis, Yves Le Traon, Márcio E. Delamaro, "Strong mutation-based test data generation using hill climbing," Proceedings of the 9th International Workshop on Search-Based Software Testing, May 2016 ,Pages 45–54, https://doi.org/10.1145/2897010.2897012
	\label{R50}
	\bibitem{RefJ}
	Behjati, R., Arisholm, E., Bedregal, M., \& Tan, C. (2019). Synthetic Test Data Generation Using Recurrent Neural Networks: A Position Paper. 2019 IEEE/ACM 7th International Workshop on Realizing Artificial Intelligence Synergies in Software Engineering (RAISE). doi:10.1109/raise.2019.00012 
	\label{R51}
	\bibitem{RefJ}
	Choi, W., Necula, G., \& Sen, K. (2013). Guided GUI testing of android apps with minimal restart and approximate learning. OOPSLA 2013, 623–640.
	\label{R52}
	\bibitem{RefJ}
	Paduraru, C. and Marius-Constantin Melemciuc. “An Automatic Test Data Generation Tool using Machine Learning.” ICSOFT (2018).
	\label{R53}
	\bibitem{RefJ}
	Sharifipour, H., Shakeri, M., \& Haghighi, H. (2018). Structural test data generation using a memetic ant colony optimization based on evolution strategies. Swarm Evol. Comput. 40, 76–91.
	\label{R54}
	\bibitem{RefJ}
	Jan Cegin, "Machine learning based test data generation for safety-critical software,Proceedings of the 28th ACM Joint Meeting on European Software Engineering Conference and Symposium on the Foundations of Software EngineeringNovember 2020 Pages 1678–1681https://doi.org/10.1145/3368089.3418538
	\label{R55}
	\bibitem{RefJ}
	Liu, P., Zhang, X., Pistoia, M., Zheng, Y., Marques, M., \& Zeng, L. (2017). Automatic Text Input Generation for Mobile Testing. ICSE 2017, 643–653.
	\label{R56}
	\bibitem{RefJ}
	E. T. Barr, M. Harman, P. McMinn, M. Shahbaz, and S. Yoo, “The oracle problem in software testing: A survey,” IEEE Trans. Softw. Eng., vol. 41, no. 5, pp. 507–525, May 2015.
	\label{R57}
	\bibitem{RefJ}
	H. Jin, Y. Wang, N. W. Chen, Z. J. Gou, and S. Wang, “Artificial neural network for automatic test oracles generation,” in Proc. Int. Conf. Comput. Sci. Softw. Eng., 2008, vol. 2, pp. 727–730.
	\label{R58}
	\bibitem{RefJ}
	F. Wang, L. W. Yao, and J. H. Wu, “Intelligent test oracle construction for reactive systems without explicit specifications,” in Proc. Int. Conf. Dependable, Auton. Secure Comput., 2011, pp. 89–96.
	\label{R59}
	\bibitem{RefJ}
	Vineeta, A. Singhal, and A. Bansal, “Generation of test oracles using neural network and decision tree model,” in Proc. Int. Conf. - Confluence Next Generation Inf. Technol. Summit, 2014, pp. 313–318.
	\label{R60}
	\bibitem{RefJ}
	Shahamiri, S. R., Kadir, W. M. N. W., Ibrahim, S., \& Hashim, S. Z. M. (2011). An automated framework for software test oracle. Inf. Softwa. Technol., 53(7), 774–788.
	\label{R61}
	
	\bibitem{RefJ}
	Braga, R., Neto, P. S., Rabêlo, R., Santiago, J.,\& Souza, M. (2018). A machine learning approach to generate test oracles. SBES 2018, 142–151.
	\label{R62}
	\bibitem{RefJ}
	W. K. Chan, J. C. F. Ho, and T. H. Tse, “Finding failures from passed test cases: Improving the pattern classification approach to the testing of mesh simplification programs,” Softw. Testing, Verification Rel., vol. 20, no. 2, pp. 89–120, 2010.
	\label{R63}
	\bibitem{RefJ}
	M. Vanmali, M. Last, and A. Kandel, “Using a neural network in the software testing process,” Int. J. Intell. Syst., vol. 17, no. 1, pp. 45–62, 2002.
	\label{R64}
	\bibitem{RefJ}
	D. Agarwal, D. E. Tamir, M. Last, and A. Kandel, “A comparative study of artificial neural networks and info-fuzzy networks as automated oracles in software testing,” IEEE Trans. Syst., Man, Cybernet., vol. 42, no. 5, pp. 1183–1193, Sep. 2012.
	\label{R65}
	\bibitem{RefJ}
	Tsimpourlas, Foivos et al. “Supervised learning over test executions as a test oracle.” Proceedings of the 36th Annual ACM Symposium on Applied Computing (2021): n. pag.
	\label{R66}
	\bibitem{RefJ}
	Spieker, Helge et al. “Reinforcement learning for automatic test case prioritization and selection in continuous integration.” Proceedings of the 26th ACM SIGSOFT International Symposium on Software Testing and Analysis (2017): n. pag.
	\label{R67}
	\bibitem{RefJ}
	Alexandre Rafael Lenz, Aurora Pozo, Silvia Regina Vergilio, Linking software testing results with a machine learning approach, Engineering Applications of Artificial Intelligence, Volume 26, Issues 5–6, 2013, Pages 1631-1640, ISSN 0952-1976, https://doi.org/10.1016/j.engappai.2013.01.008.
	\label{R68}
	\bibitem{RefJ}
	B. Busjaeger and T. Xie, “Learning for test prioritization: An industrial case study,” in Proc. ACM SIGSOFT Int. Symp. Found. Softw. Eng., 2016, pp. 975–980.
	\label{R69}
	\bibitem{RefJ}
	Wang F., Yang SC., Yang YL. (2011) Regression Testing Based on Neural Networks and Program Slicing Techniques. In: Wang Y., Li T. (eds) Practical Applications of Intelligent Systems. Advances in Intelligent and Soft Computing, vol 124. Springer, Berlin, Heidelberg. https://doi.org/10.1007/978-3-642-25658-550
	\label{R70}
	\bibitem{RefJ}
	M. Ozawa, T. Dohi, and H. Okamura, “How Do Software Metrics Affect Test Case Prioritization?,” Proc. - Int. Comput. Softw. Appl. Conf., vol. 1, pp. 245–250, 2018.
	\label{R71}
	\bibitem{RefJ}
	Lachmann, R. et al. “System-Level Test Case Prioritization Using Machine Learning.” 2016 15th IEEE International Conference on Machine Learning and Applications (ICMLA) (2016): 361-368.
	\label{R72}
	\bibitem{RefJ}
	Lachmann, R., 2018. Machine Learning-Driven Test Case Prioritization Approaches for Black-Box Software Testing. Nuremberg, Germany, European Test and Telemetry Conference, 2018.
	\label{R73}
	\bibitem{RefJ}
	Nucci, Dario Di et al. “A Test Case Prioritization Genetic Algorithm Guided by the Hypervolume Indicator.” IEEE Transactions on Software Engineering 46 (2020): 674-696.
	\label{R74}
	\bibitem{RefJ}
	Zhu, X., Zhou, B., Hou, L., Chen, J., \& Chen, L. (2008). An Experience-Based Approach for Test Execution Effort Estimation. 2008 The 9th International Conference for Young Computer Scientists. doi:10.1109/icycs.2008.53 
	\label{R75}
	\bibitem{RefJ}
	Badri, M., Badri, L., Flageol, W., \& Toure, F. (2017). Investigating the Accuracy of Test Code Size Prediction using Use Case Metrics and Machine Learning Algorithms. Proceedings of the 2017 International Conference on Machine Learning and Soft Computing - ICMLSC ’17. doi:10.1145/3036290.3036323 
	\label{R76}
	\bibitem{RefJ}
	Silva, D. G. e, Jino, M., \& Abreu, B. T. de. (2010). Machine Learning Methods and Asymmetric Cost Function to Estimate Execution Effort of Software Testing. 2010 Third International Conference on Software Testing, Verification and Validation. doi:10.1109/icst.2010.46 
	\label{R77}
	\bibitem{RefJ}
	Cheatham, T. J., Yoo, J. P., \& Wahl, N. J. (1995). Software testing. Proceedings of the 1995 ACM 23rd Annual Conference on Computer Science - CSC ’95. doi:10.1145/259526.259548 
	\label{R78}




	
	
\end{thebibliography}
\end{document}